\documentstyle[12pt]{article}
\newcommand{\be}{\begin{equation}}
\newcommand{\ee}{\end{equation}}
\newcommand{\bs}{\begin{sloppypar}}
\newcommand{\es}{\end{sloppypar}}

\newcommand{\Tr}{{\rm Tr}\,}
\newcommand{\Ker}{\mbox{Ker}\,}
\newcommand{\ind}{\mbox{ind}\,}
\newcommand{\bH}{\mbox{\bf H}}
\def\eop{ \framebox(6,6)\ {} } 

\title{New index formulas as a meromorphic generalization of the
Chern-Gauss-Bonnet theorem}

\author{
N.~V.~Borisov$^{1}$\thanks{E-mail: BORISOV@NIIF.SPB.SU},
K.~N.~Ilinski$^{2,3}$\thanks{E-mail: KNI@TH.PH.BHAM.AC.UK},
G.~V.~Kalinin$^{1}$\thanks{E-mail: KALININ@SNOOPY.NIIF.SPB.SU},
\\
{\small\it $^{1}$ Institute of
Physics, Physics Department of St-Petersburg University,}
\\
{\small\it St-Petersburg, 198904, Russian Federation.}
\\
{\small\it $^{2}$ School of Physics and Space Research, University of
Birmingham,}
\\
{\small\it Birmingham B15 2TT, United Kingdom.}
\\
{\small\it $^{3}$ Institute of Spectroscopy, Russian Academy of Sciences,}
\\
{\small\it Troitsk, Moscow region, 142092, Russian Federation.}
}
\date{ }

\begin{document}

\maketitle
\thispagestyle{empty}
\vskip -9cm
\vskip 9cm

\begin{abstract}
Laplace operators perturbed by meromorphic potential on the Riemann 
and separated  type Klein surfaces are constructed and their indices 
are calculated by two different ways.
The topological expressions for the indices are obtained from
the study of spectral properties of the operators.
Analytical expressions are provided by the Heat Kernel approach in terms of the functional integrals.
As a result two formulae connecting  characteristics of
meromorphic (real meromorphic) functions and topological properties of
Riemann (separated type Klein) surfaces are derived.
\end{abstract}

\section{Introduction}

Index theorems exhibit connection of
analytical properties of differential operators on fibre bundlers
with topological properties of these spaces. The common example is the
Atiyah-Singer index theorems~\cite{AS1} for differential operators on compact manifolds.
For the particular case of the external derivative operator
the Chern-Gauss-Bonnet theorem emerges which connects alternating sum of numbers
of harmonic form on a compact manifold (Euler characteristic of
this manifold) and the curvature integrated over the manifold.
In a similar way  the Riemann-Roch theorem and the Index theorem
for $\hat A$ genus appeared as results  of an investigation of the
Dolbeault and Dirac complexes.

New insight into the nature of index theorems was provided by E.Witten
and L.Alvarez-Gaume. It was shown that the theorems can be obtained
in a very transparent physical way treating supersymmetrical quantum
mechanical systems~\cite{Alvar,Wittens}. Following the same 
strategy several generalizations of the index theorems for noncompact 
(Euclidean at infinities) case were obtained~\cite{BMS,BI2}.
Another interesting moment produced by the supersymmetry was a new point
of view on classical
topology such as Witten's explanation of Morse Theory~\cite{Witten}.

New exciting possibility appears in the case where the index can be found via two different counting procedures. Then combining two expressions  new equalities involving more profound objects used to construct an operator can be obtained.
Such equalities no longer contain particular details of the used operator.
And Chern-Gauss-Bonnet theorem is once again an example of the approach.

In the present paper we derive two formulae connecting topological
characteristics of Riemann and separated  type Klein surfaces with integrals of some analytical
densities on them. Doing this we implicate all main points of the previous
three paragraphs. Indeed, we start with introduction of deformed
Dolbeault complex perturbed by meromorphic (real meromorphic) function on compact Riemann (separated  type Klein) surface. Then we make use of the results for indexes of the corresponding Laplace operators which were calculated in our previous
papers~\cite{BI1,BI2,DI1,DI2}. To produce correct operator picture we
need to introduce auxiliary effectively noncompact surfaces, Euclidean at infinities because of meromorphic potential (as it was
introduced at first in the Supersymmetric Scattering Theory~\cite{BMS}).
After calculation of the indices we come back to the compact
surface.
All this constitutes the next section.
In section 3 use standard Heat Kernel machinery to
obtain analytical expressions for the indices.
As a result we get our formulae.
In final section 4 we discuss the essence of the method
for deriving such kind of formulae.

Let's put down our main results -- formulae (\ref{formula11},\ref{formula12}):
\begin{equation}
\chi (M_0) + \deg \tilde D =
\int\limits_{M_0} g\frac{d\bar{z}d{z}}{2\pi i}
\exp\left(-\frac{|f^\prime|^2}{g}\right)
\left(\frac{1}{2}K-\frac{1}{g^2}|\nabla_z f^\prime|^2\right) \ ,
\label{formula11}
\end{equation}
and
\begin{equation}
\ind(f^\prime|_P: P \rightarrow \varphi(R))  = \frac{1}{2\sqrt{\pi}}
\int_P \sqrt{g} dx
\frac 1g (\nabla_z f^\prime + \overline{\nabla_z f^\prime})
\exp\left(-\frac{|f^\prime |^2}{g}\right)  \ .
\label{formula12}
\end{equation}
Here $f(z)$ in first (second) equation is a general
meromorphic (real meromorphic) function on compact Riemann surface
(dianalytic double for Klein surface of separated  type type \cite{Nat}) 
$M_0$, $g$ is an Euclidean at infinities in poles
of the function $f(z)$  K\"ahler metric on  $M_0$. 
In the first formula $K$ is a curvature of the metric and 
$\tilde D$ is a divisor of poles of the function $f(z)$, 
$\chi (M_0)$ is Euler characteristic of the surface $M_0$. In the second one
$\ind(f^\prime|_P: P \rightarrow \varphi (R))$ is
the topological index of map $f^\prime|_P$ of stationary ovals $P$ of the 
Klein surface \cite{Nat} on $\varphi(R)$ -- the image of the real axis 
under back stereographic projection $\varphi$ on Riemann sphere.

It is easy to see that Eq.(\ref{formula11}) naturally generalizes 
the Chern-Gauss-Bonnet theorem on the case of a meromorphic potential but
we find it difficult to say what the analogue of Eq.(\ref{formula12}) is.
Now we go for derivations of  Eqs.(\ref{formula11},\ref{formula12}).

\section{Operators and their indices}
In this section we do first step to derive formulae (\ref{formula11},\ref{formula12})  -- we introduce some auxiliary operators and calculate their indeces. The expressions for the indices obtained in this section will produce  the left hand sides of the
formulae. In the next section the same indices will be recalculated to obtain right hand sides of the equalities.

{\it Algebra and its index.}
An analytic index of densely defined closed operator $q: \bH_-\to\bH_+$
is defined as
$$
\ind q = dim \Ker q - dim \Ker q^* \ .
$$
Let us then introduce an auxiliary  space $\bH=\bH_+\oplus\bH_-$ with an grading involution $\tau$ and define  new self-adjoint operators

\begin{equation}
H = \left(
\begin{array}{cc} qq^* & 0 \\ 0 & q^*q
\end{array} \right)
\ ,\qquad
Q = \left(
\begin{array}{cc} 0 & q \\ q^* & 0
\end{array} \right)
\ ,\qquad
\tau = \left(
\begin{array}{cc} 1 & 0 \\ 0 & -1
\end{array} \right) \ .
\label{operSQM}
\end{equation}
These operators obey the algebra

\be
Q^* = Q \ ,\quad
\tau^* = \tau^{-1} = \tau \ ,\quad
\{Q,\tau\} = 0 \ , \quad
H = Q^{2} \ ,
\label{algSQM}
\ee
which we use intensively so far. Using the operators the index of operator $q$
can be rewritten in the following form
$$
- \ind q = dim P_{+} \Ker H - dim P_{-}\Ker H \equiv \ind (H,\tau)\ ,
$$
where $P_{\pm}$ are the projection operators for the subspaces $\bH_\pm$.

To obtain Eqs.(\ref{formula11},\ref{formula12}) we need to consider two different representation of algebra (\ref{algSQM}) and calculate the corresponding indices. The first step in this direction is the introduction of more convenient operators $2Q^\pm=Q
\pm\tau Q$ obeying the relations:

\begin{eqnarray}
(Q^-)^2 = (Q^+)^2 = 0 \ ,\quad
Q^- = (Q^+)^* \ ,\quad
\tau^* = \tau^{-1} = \tau \ ,
\nonumber \\
\{Q^\pm,\tau\} = 0 \ , \quad
H = \{Q^-,Q^+\} \ .
\label{algSQM1}
\end{eqnarray}

{\it Auxiliary Hilbert space.}
Then we can start to describe the auxiliary Hilbert space where our operators will act.
For the arbitrary genus $m$ compact Riemann surface $M_0$
and meromorphic function $f(z)$ on $M_0$ with the poles
in points $\{ z_{1},\ldots ,z_{n} \} \in M_{0} $ we
introduce the manifold $M=M_0\setminus \{z_1,\ldots ,z_n\}$ and
a K\"ahler metric $g$ on it which is Euclidean at infinity in the points
$z_1,\ldots ,z_n$ ( this means that there are open neighborhoods
$O_{R_i}$ of $z_{i}$ and diffeomorphic maps $\phi_{i}$ of
$O_{R_i}\setminus  \{z_{i}\}$ to open sets
$CB_{R_{i}}=\{u\in C : |u|>R_{i}\}$ on complex plane such that
on each $O_{R_i}$ the metric is the pullback by $\phi_{i}$
of the Euclidean metric on  $CB_{R_{i}}$; for more details see \cite{BI1}).
With such metric $g$ on $M$ the Hilbert space of the theory will be
the space of differential forms with standard scalar product:
$$
(\omega,\phi ) = \int_{M}\omega \wedge *\phi \equiv \int_{M}
\langle\omega |\phi \rangle \ ,
$$
where $*$ is the Hodge star operator.

{\it First representation of algebra (\ref{algSQM}).}
We define the operator $Q^+_c$ on smooth differential forms with
compact support $\Lambda_c(M)\equiv \bigoplus \limits^2_{k=0}
\Lambda^k_c(M)$ as deformed Dolbeault operator
\be
Q^+_c = \frac{1}{\sqrt 2}\bar{\partial }_V
= \frac{1}{\sqrt 2}(\bar\partial + V)\ ,
\label{def}
\ee
where $\bar\partial=\frac{\partial}{\partial \bar z}d\bar z\wedge $
and $V=\frac{\partial f(z)}{\partial z} dz\wedge $. Now it is easy to explain our choice of the metric.
Due to the identities as
$$
\frac{1}{\pi }\frac{\partial }{\partial\bar z}\ \frac 1z = \delta (z)
$$
operator $(Q^+_c)^2$ contains $\delta$-function terms
which have supports at the poles of $f(z)$.
The Euclidean at infinity metrics make this operator
equal to zero in the sense of Hilbert space
{\bf H} $\equiv\Lambda_2(M)=\overline{\Lambda_c(M)}$
(the closure is considered here relative to the scalar
product). The adjoint operator $Q^-_c$ and symmetric operator $Q_c$ are
defined as
$$
Q^-_c = \bigl(Q^+_c\bigl)^*\mid _{\Lambda_c(M)} \ ,\qquad
Q_c = Q^+_c + Q^-_c
$$
The operators $Q^+_c$, $Q^-_c$, $H_c=Q^+_c Q^-_c + Q^-_c Q^+_c$ and
$\tau_1 = (-1)^N$ where $N$ is the operator of the order of the
differential form give the realization of the algebra (\ref{algSQM1}).

Though the operators $Q_c,H_{c}$ give required representation they are not self-adjoint as they should be to be interpreted
 as quantum mechanical observables in the next section. To fill this lack
we define self-adjoint operator $Q$ as
\be
Q\equiv Q^{+} + Q^{-}
\label{Ch1}
\ee
with the closed operators $Q^{+},Q^{-}$
$$
Q^+\equiv \overline{Q^+_c}\ ,\qquad
Q^-\equiv \left(\overline{Q^+_c}\right)^* = (Q^+)^*
$$
on dense domains of
definition $D(Q^\pm)\ $, $\Lambda_c(M) \subset D(Q^\pm)$
and with the property ${(Q^\pm)}^2=0$.
Operators $Q_{c}$ and $H_{c}={Q_{c}}^{2}$ are essentially
self-adjoint ones and
$Q=\overline{Q_c}$, $H\equiv \overline{H_c}=Q^2$\cite{BI1}.
Perturbed Laplace operator $H$
\be
H = \frac 12 \bigl((\bar{\partial }+V)(\bar{\partial }+V)^{*}
+ (\bar{\partial }+V)^{*}(\bar{\partial }+V)\bigr)
\label{Ham1}
\ee
in local variables has the form
$$
H = H_0 + {\bar b}^* b \left(\frac{\partial}{\partial\bar z}
\overline {\frac{f^\prime}{2g}}\right) +
b^* {\bar b} \left(\frac{\partial}{\partial z}
\frac{f^\prime}{2g}\right)
\equiv H_0 + K
$$
where $H_0 = \frac 14 \Delta + \frac 1g |f^\prime|^2$,
$\Delta = 2\{\bar\partial, {\bar\partial}^*\}$ is the
usual Laplace operator on Riemannian manifold $M$ with K\"ahler metric $g$,
${\bar b}^*\equiv d\bar z\wedge ,b^{*}\equiv dz\wedge $ and $\bar b, b$
are adjoint operators for $\bar{b}^*, b^*$. The operators
$\bar{b}^*, b^*, \bar{b},b$ obey the following algebra:
$$
\{\bar{b}^*,\bar{b}\}=\{b^*,b\}=\frac 2g
\qquad \{\bar{b}^*,b\}=\{b^*,\bar b\}=0 \qquad
(\bar{b}^*)^2=(b^*)^2=(\bar{b})^2=(b)^2=0 \ .
$$
As it was shown in~\cite{BI1} that for any $\varepsilon > 0$ there
exists constant $C > 0$ such that for any $\omega \in D(H_{0})$
$$
|(K\omega ,\omega )|  \leq
\varepsilon (H_0\omega ,\omega ) + C(\omega ,\omega )
$$
This means that operator $H$ has a compact resolvent,
the space of zero-modes of perturbed  Laplace operator
has finite dimension and the following theorem takes place:

{\bf Theorem}~\cite{BI1}.
{\it Index of $H$ relative to involution $(-1)^N$
equals to the number of critical points of
function $f(z)$ (accounting in according to the degrees of
their degeneracy) with sign minus.}
\eop 

The index can be rewritten in topological terms.
Let us consider $D$ -- the
divisor of the meromorphic differential ${\partial }f$. It is
well-known~\cite{Springer} that
$$
\deg D = 2m - 2 = - \chi (M_0)
$$
where $\chi (M_0)$ is Euler characteristic of compact Riemann
surface $M_0$.  Then the number of zeros of the form ${\partial }f$ can
be expressed as a sum of Euler characteristic of Riemann surface with sign
minus and the number of poles of the differential of
the meromorphic function and
\be
\ind (H,\tau_1) = \chi (M_0) + \deg \tilde D
\label{ind1}
\ee
where $\tilde D$ is a divisor of poles of the differential of
the meromorphic function $f(z)$.
In this form the result is a pure
generalization of the results, which were obtained in~\cite{Jaffe} for
the case of $m=0$.
This fact demonstrates once more the observation that index can be
generated by either singularities or topological nontrivial situations.

The most important point to note  is that the expression (\ref{ind1})
gives LHS of the equality (\ref{formula11}).


{\it Second representation of algebra (\ref{algSQM}).}
To obtain another realization of algebra (\ref{algSQM})
we endow the Riemann surface with the Klein structure \cite{DI2}.
Compact Riemann surface is called by the Klein one if it possess
dianalytic structure and antiholomorphic involution $\tau$
\cite{Nat}. We will consider only Klein surfaces of
separated  type that is the surfaces which are
separated by stationary ovals $P = \{x=z\in M_0 :\tau z = z\}$
into disconnected parts. According to Weichold theorem
such surfaces are parametrized by $m$ genus of the surface $M_0$
and $k$, $1 \leq k \equiv m+1(mod \ 2) \leq m+1$
number of connected components of $P$ (number of stationary ovals)
i.e. two surfaces with coincident pairs $(m,k)$ are homeomorphic.
It allows us to use convenient form of an atlas and an action of the
antiholomorphic involution in it for an investigation topological
properties of such objects.

We suppose that the atlas $\left \{ (U_i, {\phi _i}) \right \}$
obeys following conditions:
$$
\phi_i(U_i)={\bar{\phi_i}}(U_i),
\quad \forall j
\quad \mbox{and} \quad
\phi_i^{-1}\left(\bar{\phi_i}(\xi)\right) =
\phi_j^{-1}\left(\bar{\phi_j}(\xi)\right),
\quad \forall i,j,\xi \in U_i\cap U_j
$$
It is easy to show that atlas possessing this property can be
constructed for surface $M_0$ with any $m,k$ and origin dianalytic
structure\cite{DI2}.
In the atlas antiholomorphic involution has simple form
\be
\forall z \in M_0 \quad z \mapsto{\bar z},\quad \bar z =
{\bar\phi}^{-1}\left(\phi(z)\right)
\label{CC}
\ee

To make other structures consistent with the Klein one
we impose the restrictions:
\begin{enumerate}
\item The function $f(z)$ is real meromorphic
\be
f(\bar z)=\overline {f(z)}\ ;
\label{real-m}
\ee
\item The
K\"ahler metric $g$ is symmetrical with respect to the
antiholomorphic involution
$$
g(z,\bar z) = g(\bar z,z) \ .
$$
\end{enumerate}
Condition (\ref{real-m}) allows us to introduce surface
$M=M_0\setminus \{z_1,\ldots ,z_n\}$ and an Euclidean at infinity
K\"ahler metric $g$ similar to the previous consideration
keeping all structures.

The new symmetry generated by the Klein structure supply a new realization
of algebra (\ref{algSQM}). Let us consider new operator
\be
Q = \frac 12 \left[\bar\partial - \partial + V - \bar V + h.c.
\right]\ ,
\label{Ch2}
\ee
where $\partial= \frac{\partial}{\partial z} dz\wedge$,
$\bar V = \overline{\frac{\partial f}{\partial z}} d\bar z\wedge$.
In contrast to  operator~(\ref{Ch1}) the new operator $Q$ anticommutes
not only with involution $\tau_1 = (-1)^N$ but also with
"complex conjugation" involution $\tau_2$:
$$
\tau_2 \  :\quad z \rightarrow {\bar
z}, \quad {\bar z} \rightarrow z, \quad dz \rightarrow d{\bar z}, \quad
d{\bar z} \rightarrow dz,
$$
generated by transformation~(\ref{CC}).
Operators $Q,H,\tau_2$ forms algebra~(\ref{algSQM}).

{\bf Theorem}~\cite{DI2}.
{\it
\be
\ind (H,\tau_2) = -\ind(f^\prime|_P: P \rightarrow \varphi(R))
\label{ind2}
\ee
where $\ind(f^\prime|_P: P \rightarrow \varphi(R))$ is the index of the map
$f^\prime|_P$ which reflects stationary ovals $P$ on $\varphi(R)$,
$\varphi(R)$ is the image of the real axis under back
stereographic projection $\varphi$ on Riemann sphere,
i.e. the sum of the indices of the map $f^\prime$ of connected
components of $P$.}
\eop\\
The result of the last theorem gives us the LHS of equality (\ref{formula12}).

\section{Supersymmetric interpretation}

Now we go for the RHS of Eqs.(\ref{formula11},\ref{formula12}). 
To obtain the required expressions we will calculate 
$\ind (H,\tau_1)$, $\ind (H,\tau_2)$ in the Heat Kernel approach. 
More precisely, we will follow Refs.\cite{Alvar} and
make use supersymmetric quantum mechanical analogy.

{\it Supersymmetric quantum mechanics.}
Physicists would refer to algebra (\ref{algSQM}) or (\ref{algSQM1}) as the
algebra of Supersymmetric Quantum Mechanics with the involution $\tau$
\cite{Wittens} (or, for a few supersymmetric involutions, algebra 
of Generalized Supersymmetric quantum mechanics \cite{BIU}).  The works 
by E.Witten \cite{Wittens,Witten}, A.Jaffe \cite{Jaffe,Jaffe1}, 
L.Alvarez-Gaume \cite{Alvar} and others prove that the 
Supersymmetric Quantum Mechanics and Supersymmetric Quantum Field Theory 
are powerful tools to connect and splice geometrical and analytical 
substances.  The Witten's approach to the deriving of Morse's 
inequalities, supersymmetric proof of Index Theorems, the construction of 
infinitedimensional analysis on the base of Supersymmetric Quantum Field 
Theory are some examples of such connections.

The model which we will concentrate on is so-called Meromorphic
Quantum mechanics. It was studied in a number of papers
\cite{Arai,BI1,BI2,Cecotti,DI1,DI2,IKM,Jaffe,Jaffe1,Klimek}
(the corresponding field theory was studied in
\cite{Cecotti, Jaffe1, Klimek1}). In certain sense, the opposite 
case was investigated in \cite{Moroz} where the index was defined and
calculated for the non-Fredholm case of Aharonov-Bohm potential 
in the Dirac operator: we removed this difficulty from the theory 
using Euclidean at infinity metric.

In this context
the operators $Q, Q^\pm$ and $H$ are called the supercharges
and Supersymmetric Hamiltonian. The grading operator $\tau$ defines a separation of the Hilbert space of the theory into bosonic(fermionic) subspaces
{\bf H}$_{\pm}$. Then the corresponding
index is called Witten index of the supersymmetric Hamiltonian $H$:
$$
\Delta _{W}(H,\tau ) = dim P_{+} \Ker H -
dim P_{-}\Ker H \ .
$$
Due to Supersymmetry reflected by the algebra (\ref{algSQM}), 
bosonic and fermionic subspaces 
{\bf H}$_{\pm}(E_k)$ of an eigenspace {\bf H}$(E_k)$
of Supersymmetric Hamiltonian $H$ have equal dimensions
for any eigenvalue $E_k>0$.
This property leads to topological stability of the
Witten index and gives possibility to write
Supertrace representation for the Witten index:
\begin{equation}
\Delta_W(H,\tau ) = \Tr \left( \tau e^{-\beta H} \right)
\ ,\qquad\forall \beta > 0 \ ,
\label{str}
\end{equation}
which is a starting point in the Heat Kernel approach to the index calculation.
Because of our quantum mechanical interpretation we can consider
the  RHS of~(\ref{str}) as a
partition function for the system with the Hamiltonian $H$ at finite temperature $1/\beta$ "twisted" by the involution $\tau$. To evaluate the partition function
the corresponding function integral will be used.

{\it RHS of Equality (\ref{formula11}).}
The quantum mechanical system with our Hamiltonian $H$
can be obtained by the quantization of a classical system with the Lagrangian
$$
\begin{array}{rl}
\tilde L = & g \mid \dot z  \mid ^{2} - {i} g (
\dot { \bar \psi _{1}} \psi _{1} + \dot { \bar \psi _{2}} \psi _{2} )-
\\ & - i g _{z} \dot z \bar \psi _{2} \psi _{2} - i g _{ \bar z }
\dot {\bar z }  \bar \psi _{1} \psi _{1} +
\frac{g^2}{2} K \bar \psi _{1} \psi _{1} \bar \psi _{2} \psi _{2} -
\\ & - \nabla _{z} f^\prime \bar \psi _{2}\psi _{1} - \nabla _{\bar z}
\bar f_{\bar z} \bar \psi _{1}\psi _{2} -
\frac {\mid f^\prime \mid ^{2}} {g}
\end{array}
$$
with the Gauss curvature of surface $M$
$$
K=-\frac 2g \frac{\partial^2}{\partial z \partial \bar z } ln(g)=
2(\frac{g_{z}g_{\bar z}}{g^3}-\frac{g_{z\bar z}}{g^2})
$$
and covariant derivatives
$$
\nabla_{z}=\frac{\partial}{\partial z}-\frac{g_{z}}{g} \qquad
\nabla_{\bar z}=\frac{\partial}{\partial \bar z}-\frac{g_{\bar z}}{g} \ .
$$
Since the Lagrangian is known the functional integral representation for
the matrix element of the statistical operator follows immediately~\cite{Popov}:
\begin{eqnarray*}
 & & \langle \bar{\psi}_2, \bar{\psi}_1, z^\prime, \bar{z}^\prime|
\tau_1 e^{-\beta H} |z, \bar{z}, \psi_1, \psi_2 \rangle = \\
 & & \int \prod\limits_{t\in [0,\beta]}
 \frac{gd\bar z(t) dz(t)}{2\pi i}
 d\bar{\psi}_1(t)d\psi_1(t)d\bar{\psi}_2(t)d\psi_2(t)
 \exp\left(- \int \limits_{0}^{\beta}L dt +
 \sum_{j=1}^{2}\bar{\psi_j}(0) \psi_j(0) \right)
\end{eqnarray*}
where
\begin{eqnarray}
L & = & g | \dot z |^{2} - (
\dot { \bar \psi _{1}} \psi _{1} + \dot { \bar \psi _{2}} \psi _{2} ) -
\frac{g_{z}}{g} \dot z \bar \psi _{2} \psi _{2} -
\frac{g_{\bar z}}{g} \dot {\bar z}  \bar \psi_{1} \psi_{1} -
\nonumber\\
 & - & \frac{1}{2} K \bar \psi _{1} \psi _{1} \bar \psi _{2} \psi _{2} +
\frac{1}{g}\nabla _{z} f^\prime \bar \psi _{2}\psi _{1}+
\frac{1}{g}\nabla _{\bar z}
\bar f_{\bar z} \bar \psi _{1}\psi _{2} +
\frac {\mid f^\prime \mid ^{2}} {g} \ ;
\label{lag}
\end{eqnarray}
$z(t), \bar{z}(t)$ are the complex functions on $[0, \beta]$
with the boundary conditions
$$
z(0) = z \ ,\quad \bar{z}(0) = \bar{z} \ ,\quad
z(\beta) = z^\prime \ ,\quad \bar{z}(\beta) = \bar{z}^{\prime} \ .
$$
Functions $\bar{\psi}_{1,2}(t), \psi_{1,2}(t)$ are grassmannian functions
on $[0, \beta]$ which due to the involution $\tau_1$
obey the following boundary conditions
$$
\psi_{1,2}(0) = \psi_{1,2} \ ,\quad
\bar{\psi}_{1,2}(\beta) = - \bar{\psi}_{1,2} \ .
$$
Using the Supertrace representation (\ref{str}) for the index and the formula
for trace
\begin{eqnarray}
\Tr \left( \tau e^{-\beta H}\right)& = &
\int \frac{gd\bar{z}dz}{2\pi i}
d\bar{\psi}_1d\psi_1d\bar{\psi}_2d\psi_{2}
\exp\left(-\sum_{j=1}^{2}\bar{\psi_j} \psi_j \right)
\times
\nonumber \\
& & \times \langle -\bar{\psi}_2, -\bar{\psi}_1, z, \bar z|
\tau e^{-\beta H} |z, \bar{z}, {\psi}_1, {\psi}_2 \rangle \ ,
\label{trace}
\end{eqnarray}
we get functional integral representation for Witten index:
\be
\Delta_{W}(H,\tau_1) =
\int \prod\limits_{t\in \beta} \frac{gd\bar{z}(t)dz(t)}{2\pi i}
d\bar{\psi}_1(t)d\psi_1(t)d\bar{\psi}_2(t)d\psi_{2}(t)
\exp\left(- \int \limits_{0}^{\beta}L dt \right)
\label{int1}
\ee
with periodic on $[0,\beta]$ boundary condition for the fields
$z(t), \bar{z}(t), \bar{\psi}_{1,2}(t), \psi_{1,2}(t)$.

The index does not depend on the value of $\beta$ so we can treat it
as $\beta$ independent term in functional integral~(\ref{int1})
in the limit $\beta\to 0$ \cite{Alvar}.
To this end we expand
$z(t), \bar{z}(t), \bar{\psi}_{1,2}(t), \psi_{1,2}(t) $
in  Fourier series with frequencies $\frac{2\pi n}{\beta}$
and split the integral in two parts. First is the integral
over zero modes coefficients
$z_0, \bar{z}_0, \bar{\psi}_{1,2}^0, \psi_{1,2}^0$,
second is the integral over the remaining Fourier coefficients.
The latter can be evaluated by saddle point method as a series on $\beta$.
The first term of the series is the fraction of bosonic and fermionic
determinants which is equal to $1$ due to the Supersymmetry.
This is the only term we need because
the other ones contain positive powers of the parameter $\beta$.
The former (after rescaling variables
$z=z_0/\beta^{\frac 12}$, $\bar z=\bar z_0/\beta^{\frac 12}$,
$\bar \psi_{1,2}=\bar \psi_{1,2}^0/\beta^{\frac 14}$
$\psi_{1,2}=\psi_{1,2}^0/\beta^{\frac 14}$)
has the form

\begin{eqnarray*}
\Delta_{W}(H,\tau_1)& = & \int
\frac{gd\bar zdz}{2\pi i}
d\bar{\psi_1}d\psi_1d\bar{\psi_2}d\psi_{2} \exp(
\frac 12 K \bar\psi_{1} \psi_{1} \bar\psi_{2} \psi_{2} -  \\
& & - \frac{\sqrt{\beta}}{g} \nabla_{z} f^\prime \bar \psi_{2}\psi_{1} -
\frac{\sqrt{\beta}}{g} \nabla_{\bar z} \bar {f^\prime} \bar\psi_1 \psi_2
- \frac {\beta |f^\prime |^{2}}{g})
\end{eqnarray*}
and after the integration over grassmannian variables
$\bar\psi_{1,2}, \psi_{1,2}$
becomes the integral over the manifold $M$:

\begin{equation}
\Delta_W (H,\tau_1) = \int\limits_M
\frac{gd\bar{z}dz}{2\pi i}
\exp\left(-\frac{\beta |f^\prime|^2}{g}\right)
\left(\frac 12 K-\frac{\beta}{g^2}|\nabla_z f^\prime|^2\right)
\ .
\label{otbeta1}
\end{equation}
By a simple and straightforward calculation it is easy to check that the RHS
is independent on $\beta$:

\begin{eqnarray*}
\frac{d}{d\beta}\int\limits_{M}
\frac{gd\bar{z}d{z}}{2\pi i}
\exp\left(-\frac{\beta |f^\prime|^2}{g}\right)
\left(\frac{1}{2}K-\frac{\beta}{g^2}|\nabla_z f^\prime|^2\right)
& = &\\ =
\frac{1}{\beta}\int\limits_M \frac{d\bar{z}d{z}}{2\pi i}
\frac{\partial^2}{\partial z\partial\bar z}
\exp\left(-\frac{\beta|f^\prime|^2}{g}\right)& = & 0
\end{eqnarray*}
Similarly we can show the
stability property of the RHS of (\ref{otbeta1})
under a smooth deformation of the metric. For the simplicity
we choose the deformation in the form $g(t)=(1+th)g$, then
the following chain of equalities are true:
\begin{eqnarray*}
\frac{d}{dt}\int\limits_{M}
\frac{g(t)d\bar{z}dz}{2\pi i} \exp
\left(-\frac{|f^\prime|^2}{g(t)}\right)
\left(\frac{1}{2}K(g(t))-\frac{|\nabla_z f^\prime|^2}{(g(t))^2}\right)
\left.\over\right|_{t=0} = &\\
= - \int\limits_{M} \frac{d\bar{z}d{z}}{2\pi i}
\frac{\partial^2}{\partial z\partial\bar z}
\left(h\exp\left(-\frac{|f^\prime|^2}{g}\right)\right) = &0 \ .
\end{eqnarray*}

Summarizing, we get an analytic expression for the Witten index
in the form of an integral over the manifold.
Due to the existence of the cutting exponent in (\ref{otbeta1})
we can make a limiting procedure for the
integration over $M$ with the Euclidian at infinity metric
and substitute it by the integration over $M_0$.
Comparing this expression with the topological expression for
index (\ref{ind1}) we get the formula which
connects Euler characteristic $\chi (M_0)$
of compact Riemann surface $M_0$ and a divisor of poles
$\tilde D$ of the differential of a meromorphic function
$f(z)$ with an integral over the surface of the density
dependent on function $f$ and the Euclidean at infinities K\"ahler
metric $g$:
\begin{equation}
\chi (M_0) + \deg \tilde D =
\int\limits_{M_0} g\frac{d\bar{z}d{z}}{2\pi i}
\exp\left(-\frac{|f^\prime|^2}{g}\right)
\left(\frac 12 K-\frac{1}{g^2}|\nabla_z f^\prime|^2\right)
\label{formula1}
\end{equation}
This gives us the formula (\ref{formula11}).

{\it RHS of Equality (\ref{formula12}).}
In this subsection we consider Witten index for the Klein representation
of the supersymmetry algebra.
Hence we can use the functional integral representation for a
matrix element of the partition function "twisted" by the
involution $\tau_2$. It has the following functional integral
representation~\cite{Popov}:

\begin{eqnarray*}
 & & \langle \bar{\psi}_2, \bar{\psi}_1, {z}^{\prime}, \bar{z}^{\prime}|
\tau_2 e^{-\beta H} |z, \bar{z}, {\psi}_1, {\psi}_2 \rangle = \\
 & & \int \prod\limits_{t\in [0,\beta]} \frac{gdz(t)d\bar{z}(t)}{2\pi i}
d\bar{\psi}_1(t)d\psi_1(t)d\bar{\psi}_2(t)d\psi_{2}(t)
\exp\left(- \int \limits_0^\beta L dt +
\sum_{j=1}^{2}\bar{\psi_j}(0) \psi_j(0) \right)
\end{eqnarray*}
where $L$ is given by Eq.(\ref{lag}) and
boundary condition for the field of integration
$z(t), \bar{z}(t)$, $\bar{\psi}_{1,2}(t), \psi_{1,2}(t)$
on $[0, \beta]$ are
\begin{eqnarray*}
z(0) = z \ ,\quad \bar{z}(0) = \bar{z} \ ,\quad
z(\beta) = \bar{z}^{\prime } \ ,\quad \bar{z}(\beta) = z^\prime \ ,\\
\psi_{1,2}(0) = \psi_{1,2} \ ,\quad
\bar{\psi}_{1,2}(\beta) = \bar{\psi}_{2,1} \ .
\end{eqnarray*}

Using Supertrace representation (\ref{str}) and the formula
for the trace (\ref{trace}) we get the functional integral representation
for the index $\Delta_{W}(H,\tau_{2})$:
\be
\Delta_{W}(H,\tau_{2}) =
\int \prod\limits_{t\in \beta} \frac{gd\bar{z}(t)dz(t)}{2\pi i}
d\bar{\psi}_1(t)d\psi_1(t)d\bar{\psi}_2(t)d\psi_{2}(t)
\exp\left(- \int \limits_{0}^{\beta}L dt \right) \ .
\label{int2}
\ee
This integral is provided with the boundary conditions for the fields:
\be
z(0) = \bar{z}(\beta) \ ,\quad \bar{z}(0) = z(\beta) \ ,\quad
\bar{\psi}_{1,2}(0) = -\bar{\psi}_{2,1}(\beta) \ ,\quad
{\psi}_{1,2}(0) = - {\psi}_{2,1}(\beta) \ .
\label{*3}
\ee
In terms of real variables
$$
x^1= \frac{z+\bar z}{2} \ ,\quad
x^2= \frac{z-\bar z}{2i} \ ,\quad
$$
$$
\bar\xi_1= \frac{\bar\psi_1+\bar\psi_2}{2} \ ,\quad
\bar\xi_2= \frac{\bar\psi_1-\bar\psi_2}{-2i} \ ,\quad
\xi_1 = \frac{\psi_1+\psi_2}{2} \ ,\quad
\xi_2 = \frac{\psi_1-\psi_2}{2i} \ .
$$
the boundary conditions~(\ref{*3}) have the common form:
$$
x^k(0) = (-1)^{k+1}x^k(\beta) \ ,\quad
\xi_k(0) = (-1)^k\xi_k(\beta) \ ,\quad
\bar{\xi}_k(0) = (-1)^k\bar\xi_k(\beta) \ ,
$$
i.e. they are periodic on $[0,\beta]$ for bosonic
coordinate $x^1$ and fermionic coordinates ${\psi}_2, \bar{\psi}_2$
and antiperiodic for $x^2, {\psi}_1, \bar{\psi}_1$.

We calculate Witten index $\Delta_W(H,\tau_2)$ again as
$\beta$ independent term in the functional integral~(\ref{int2})
in the limit $\beta\to 0$. According to the boundary conditions we
expand fields $x^1(t), \bar\xi_2(t), \xi_2(t)$ and
$x^2(t), \bar\xi_1(t), \xi_1(t)$
in a Fourier series with the frequencies $\frac{2\pi n}{\beta}$
and $\frac{(2n+1)\pi}{\beta}$ respectively.
Fourier series for the functions $x^2(t), \bar\xi_1(t), \xi_1(t)$
do not contain zero modes therefore in the  limit $\beta\to 0$
only the stationary paths $x^2(t)=0, \psi_1(t)=0, \bar\psi_1(t)=0$
contribute in the integral over these fields. This is why bosonic
variables ($x^1, x^2$) lie on the stationary ovals $P$
of the involution $z\to\bar z$.

For integration over $x^1(t), \bar\xi_2(t), \xi_2(t)$
we again split the integral in two parts over zero modes coefficients
$x^1_0, \bar\xi_2^0, \xi_2^0$ and over the remaining Fourier coefficients.
The latter can be evaluated by saddle point method as a series on $\beta$.
The first term of the series is the fraction of bosonic and fermionic
determinants which is again equal to $1$ due to Supersymmetry.
This is the only term we need because
the other ones contain positive powers of $\beta$.
The resulting integral (after rescaling variables
$x^1 = x_0^1/\beta^{\frac 12},
\bar\xi_2 = \bar\xi_2^0/\beta^{\frac 14},
\xi_2 = \xi_2^0/\beta^{\frac 14}$)
has the form
$$
\Delta_W (H,\tau_2) =
\frac{1}{2\sqrt{\pi}}\int dx^1 d\bar\xi_2 d\xi_2
\frac{1}{\sqrt{g} }
\exp\left(-\frac{|f^\prime|^2}{g}\right)
\exp\left({\bar\xi_2 \xi_2 (\nabla_z f^\prime +
\overline{\nabla_z f^\prime})} \right) \ ,
$$
and after the integration over grassmannian variables
$\bar\xi_2, \xi_2$ we get
\be
\Delta_W (H,\tau_2) = \frac{-1}{2\sqrt{\pi}}
\int\limits_P \sqrt{g} dx \frac{\sqrt\beta}{g}
(\nabla_z f^\prime + \overline{\nabla_z f^\prime})
\exp\left(-\frac{\beta|f^\prime|^2}{g}\right) \ .
\label{otbeta2}
\ee
Here the integration is taken over all stationary ovals determined by the equation
$x^2 = 0$ in the atlas consisted of maps which are
invariant with respect to complex conjugation described by
transformation $x^2 \mapsto -x^2$.

It is easy to see that the last expression is independent on $\beta$
\begin{eqnarray*}
\frac{d}{d\beta}
\int\limits_P \sqrt{g} dx
\frac{\sqrt\beta}{g} (\nabla_z f^\prime + \overline{\nabla_z f^\prime})
\exp\left(-\frac{\beta|f^\prime|^2}{g}\right)
&=&\\=
\int\limits_P dx \frac{\partial}{\partial x} \left(
\frac{f^\prime}{\sqrt{g\beta}}
\exp\left(-\frac{\beta|f^\prime|^2}{g}\right)\right) &= &0
\end{eqnarray*}
and a smooth deformation of the metric $g(t)=(1+th)g$
\begin{eqnarray*}
\frac{d}{dt}
\int\limits_P \sqrt{g} dx
\exp\left(-\frac{|f^\prime|^2}{g}\right)
\frac 1g (\nabla_z f^\prime + \overline{\nabla_z f^\prime})
\left.\over\right|_{t=0} &= &\\
= - \int\limits_P dx \frac{\partial}{\partial x} \left(
h\frac{f^\prime}{\sqrt{g}}
\exp\left(-\frac{|f^\prime|^2}{g}\right)\right) &= &0 \ .
\end{eqnarray*}
These equalities prove the topological invariance of the index in a straightforward way.

Summarizing, the equation~(\ref{otbeta2}) gives analytic expression
for Witten index. Now we can come back to compact Klein
surface $M_0$  due to the existence of the
cutting exponent. Using the topological expression for the
index~(\ref{ind2}) we obtain formula which gives analytic expression
for the topological index of map $f^\prime|_P$ of stationary ovals $P$ on
$\varphi(R)$ the image of the real axis under the back
stereographic projection $\varphi$ on the Riemann sphere:
\begin{equation}
\ind(f^\prime|_P: P \rightarrow \varphi(R))  = \frac{1}{2\sqrt{\pi}}
\int \sqrt{g} dx
\frac 1g (\nabla_z f^\prime + \overline{\nabla_z f^\prime})
\exp\left(-\frac{|f^\prime |^2}{g}\right)  \ .
\label{formula2}
\end{equation}
This completes our derivation of formula (\ref{formula12}).

\section{Conclusion and perspectives}

In the paper we derived two formulae connecting topological and analytical
characteristics of arbitrary meromorphic function on Riemann and
Klein surfaces. We did this using well-known and well-developed machinery
of the supersymmetric quantum mechanics. Namely, we calculated
the Witten index by means of the operator theory and the functional
integration. The only new entry was an implication of meromorphic
superpotential and the Euclidean at infinity metric.

However we would like to note that there are other way to go in the same direction. Indeed, it is possible to consider known models but use other quantum mechanical topological invariants for auxiliary quantum systems. Inspired by successes of the supersy

mmetric quantum mechanics with
the Witten index as a main tool other characteristics of supersymmetric quantum mechanical systems can be considered to  produce (possibly) a  new outcome.
To proceed in this way we need to find quantum mechanical spectral
quantities such that:
\begin{enumerate}
\item They are topologically stable; This stability for the quantities provided by the fact that they only depends on vacuum
state properties since the contributions of superpartners for nonzero energy are vanishing.
\item There is  a machinery of Quantum Field Theory methods for calculation of
such quantities (as it is for the Witten index which can be realized as
a partition function
"twisted" by the supersymmetric involution (\ref{str}) and so far can be
presented in the form of a functional integral).
\end{enumerate}
In addition to the Witten index, which is used in the  paper, there are
other topologically stable indices in the supersymmetric frameworks: 
Supersymmetric Scattering Index
in the Supersymmetric Scattering Theory \cite{BMS},  GSQM-indices
\cite{BIU} in Generalized Supersymmetric Quantum Mechanics connected with
q- deformation of Extend supersymmetric Quantum Mechanics \cite{IU}
and Supersymmetric Berry Index for cyclic adiabatic evolution of
supersymmetric system \cite{IKM}. Various parasupersymmetric indices
may be introduced as well but as far as we know  there is no geometrically 
interesting examples of systems in question.
They all can be considered as candidates for  derivation of new
index theorems.

{\it Acknowledgement.} We thank V.Matveev and S.Natanzon for 
helpful discussions.
This work was supported 
partially (K.I., G.K.) by
the Russian Fund of Fundamental Investigations, Grant N 95-01-00548 
and 
partially (N.B., G.K.) by
the Russian Fund of Fundamental Investigations, Grant N 96-01-00535,
GRACENAS Grant No. 95-0.6-49

\end{document}